\documentclass[prb,aps,amsmath,amssymb,showpacs,twocolumn]{revtex4-1}
\usepackage{amssymb,graphicx,float,dcolumn,bm,subfigure,color}

\begin{document}

\title{Role of exciton screening in the 7/3 fractional quantum Hall effect}
\author{Ajit C. Balram,$^1$ Ying-Hai Wu,$^1$ G. J. Sreejith,$^{1,2}$ Arkadiusz W\'ojs,$^3$ and Jainendra K. Jain$^1$}
\affiliation{
   $^{1}$Department of Physics, 
   104 Davey Lab, 
   Pennsylvania State University, 
   University Park PA, 16802}
\affiliation{
	$^{2}$NORDITA, 
	Roslagstullsbacken 23, 
	10691 Stockholm, 
	Sweden}
\affiliation{
   $^{3}$Institute of Physics, 
   Wroclaw University of Technology,
   50-370 Wroclaw, Poland}

\date{\today}

\begin{abstract} 
The excitations of the 7/3 fractional Hall state, one of the most prominent states in the second Landau level, are not understood. 
We study the effect of screening by composite fermion excitons and find that it causes a strong renormalization at 7/3, thanks to a relatively small exciton gap and a relatively large residual interaction between composite fermions. The excitations of the 7/3 state are to be viewed as composite fermions dressed by a large exciton cloud. Their wide extent has implications for experiments as well as for analysis of finite system exact diagonalization studies.
\end{abstract}

\pacs{73.43.-f, 05.30.Pr, 71.10.Pm}

\maketitle

While the fractional quantum Hall effect (FQHE) at filling factor 1/3 was the first one to be understood,\cite{Laughlin83} its counterpart in the second Landau level (LL), namely the 7/3 state, has remained a puzzle. In particular, no understanding exists of the excitations of the 7/3 state, which are seen, in finite-size exact-diagonalization studies, to be strikingly different from those at 1/3. Remembering that excitations are an integral part of a theory and a fundamental manifestation of the topological structure of a FQHE state, this raises the question whether 7/3 and 1/3 states have the same underlying physics. A proper understanding of the 7/3 state has become all the more important in view of the recent measurements of its shot noise,\cite{Dolev,Gross} local electrochemical potential,\cite{Venkatachalam} Aharonov Bohm interference,\cite{Willett,Kang} tilted field transport,\cite{Eisenstein} and spin polarization.\cite{Wurstbauer}

To gain insight into this issue, we study the role of screening by composite fermion (CF) excitons\cite{Jain89,Scarola,Girvin} at both 1/3 and 7/3, and find that it causes a substantial renormalization of the quasiparticle and quasihole at 7/3. The actual excitations of the 7/3 state should therefore be viewed as composite fermions carrying a large  exciton cloud. This has relevance to experimental studies of the 7/3 FQHE, and at the same time shows how one can rationalize the differences between 7/3 and 1/3 seen in finite system studies within a common framework.

\begin{figure}
\includegraphics[width=0.45\textwidth]{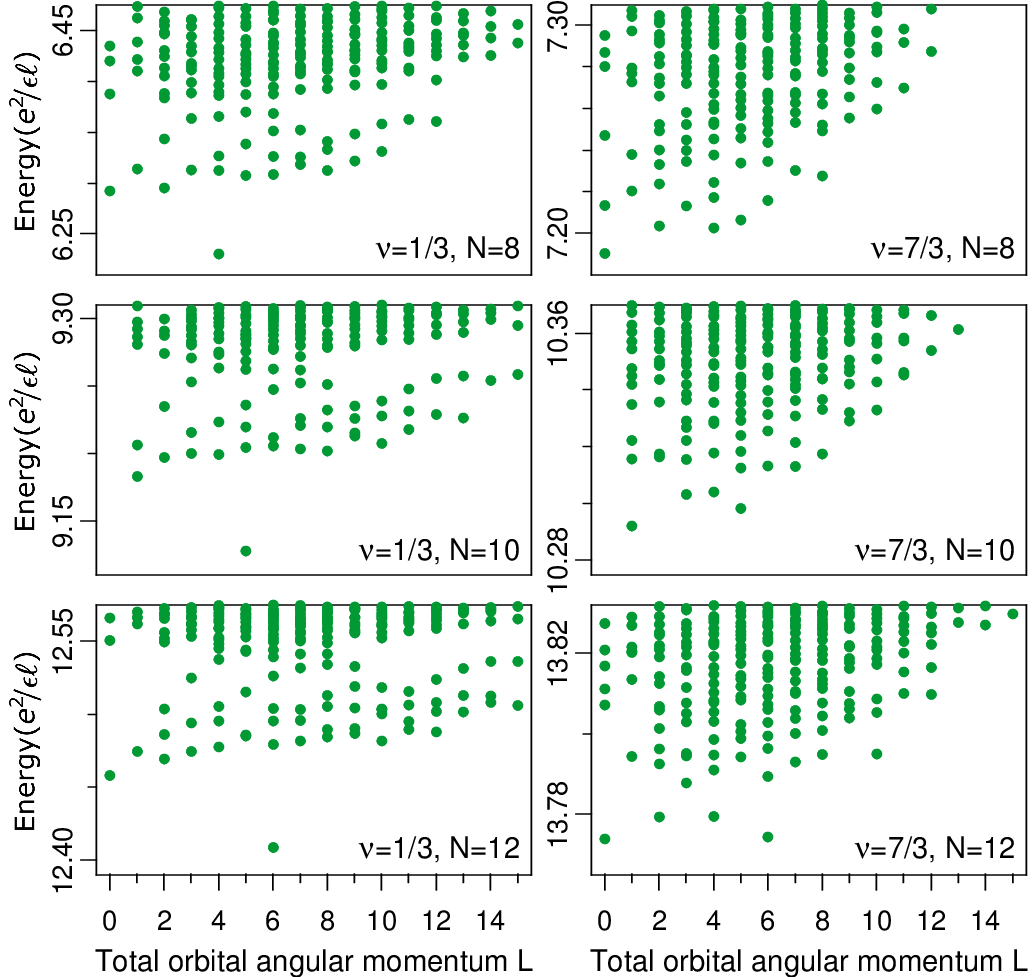}
\caption{Comparing quasiholes at 1/3 and 7/3. Left panels show the exact Coulomb spectra in the spherical geometry for total flux $2Q=3N-2$ in the lowest LL. The right panels show corresponding spectra at 7/3. $N$ is the number of electrons, and the energy is quoted in units of $e^2/\epsilon \ell$, where $\ell=\sqrt{\hbar c/eB}$ is the magnetic length and $\epsilon$ is the dielectric constant of the host material. The lowest energy state at 1/3 at total orbital angular momentum $L=N/2$ is identified as the 1/3 QH. The spectrum at 7/3 has a very different structure.}
\label{fig:quasihole}
\end{figure}

\begin{table}[b]
\begin{tabular} {|c | c c c c c c c c |}\hline
$N$ & 8    &  9  &  10  &  11  &  12  &  13  &  14  &  15  \\ \hline
ground state & 0.39 & 0.29 & 0.37 & 0.59 & 0.33 & 0.39 & 0.44 & 0.38 \\
$L=N/2$ QP & 0.18 & 0.17 & 0.23 & 0.13 & 0.18 & 0.24 & 0.12 & 0.17 \\
$L=N/2$ QH & 0.39 & 0.31 & 0.33 & 0.41 & 0.23 & 0.34 & 0.33 & 0.27 \\ \hline
\end{tabular}
\caption{Comparison of the 1/3 and 7/3 FQHE. Squared overlaps are given for the corresponding ground states, quasiparticles (QPs) and quasiholes (QHs), all evaluated using the exact Coulomb eigenstates. $N$ is the total number of particles and $L$ is the total orbital angular momentum. At 7/3 the lowest energy QP and QH states occur at $L\neq N/2$; we have taken the lowest energy state in the $L=N/2$ sector to evaluate the overlaps. The dimension of the Fock space for $N=15$ at $2Q=43$ is $> 2.2\times 10^9$.  For the ground state, the overlaps for up to $N=9$ were obtained previously.\cite{Ambrumenil_Reynolds_88}}
\label{over}
\end{table}

{\em Exact diagonalization}: 
We begin by quoting exact diagonalization results. The spherical geometry\cite{Haldane_83} is used for all computations below, which considers $N$ electrons on the surface of a sphere, with a total flux $2Q$ (measured in units of the ``flux quantum" $\phi_0=hc/e$) passing through the surface. It is assumed that the system is fully polarized, the width is zero, and LL mixing is negligible. The phrase ``7/3 state" refers to 1/3 state in the second LL. It differs from the 1/3 state in the lowest LL (LLL) because the Coulomb interaction has different matrix elements in the two LLs. The eigenstates are labeled by the total orbital angular momentum $L$. 

The Laughlin incompressible state corresponds to $2Q=3N-3$. Here, the Coulomb ground states at both 1/3 and 7/3, obtained from exact diagonalization, occur at $L=0$, although the overlap between the two, given in Table~\ref{over}, is not very high. The quasihole (QH) and quasiparticle (QP) are obtained by adding or removing a flux quantum, i.e. at $2Q=3N-2$ and $2Q=3N-4$, respectively.\cite{comm} For 1/3, the lowest state at both these values of $2Q$ occurs at $L=N/2$ (see Fig.~\ref{fig:quasihole} for $2Q=3N-2$), which is well separated from the continuum of other excitations; this state is identified with the QP or QH. The structure at 7/3 is strikingly different. As seen in Fig.~\ref{fig:quasihole} for $2Q=3N-2$, no single state can be identified as a QH, and the lowest state occurs at $L\neq N/2$ for up to $N=15$. The same is true for the QP. Furthermore, the lowest state in the $L=N/2$ sector is very different from the 1/3 QP / QH, as seen from the overlaps in Table.~\ref{over}.

\begin{figure}
\includegraphics[width=0.45\textwidth]{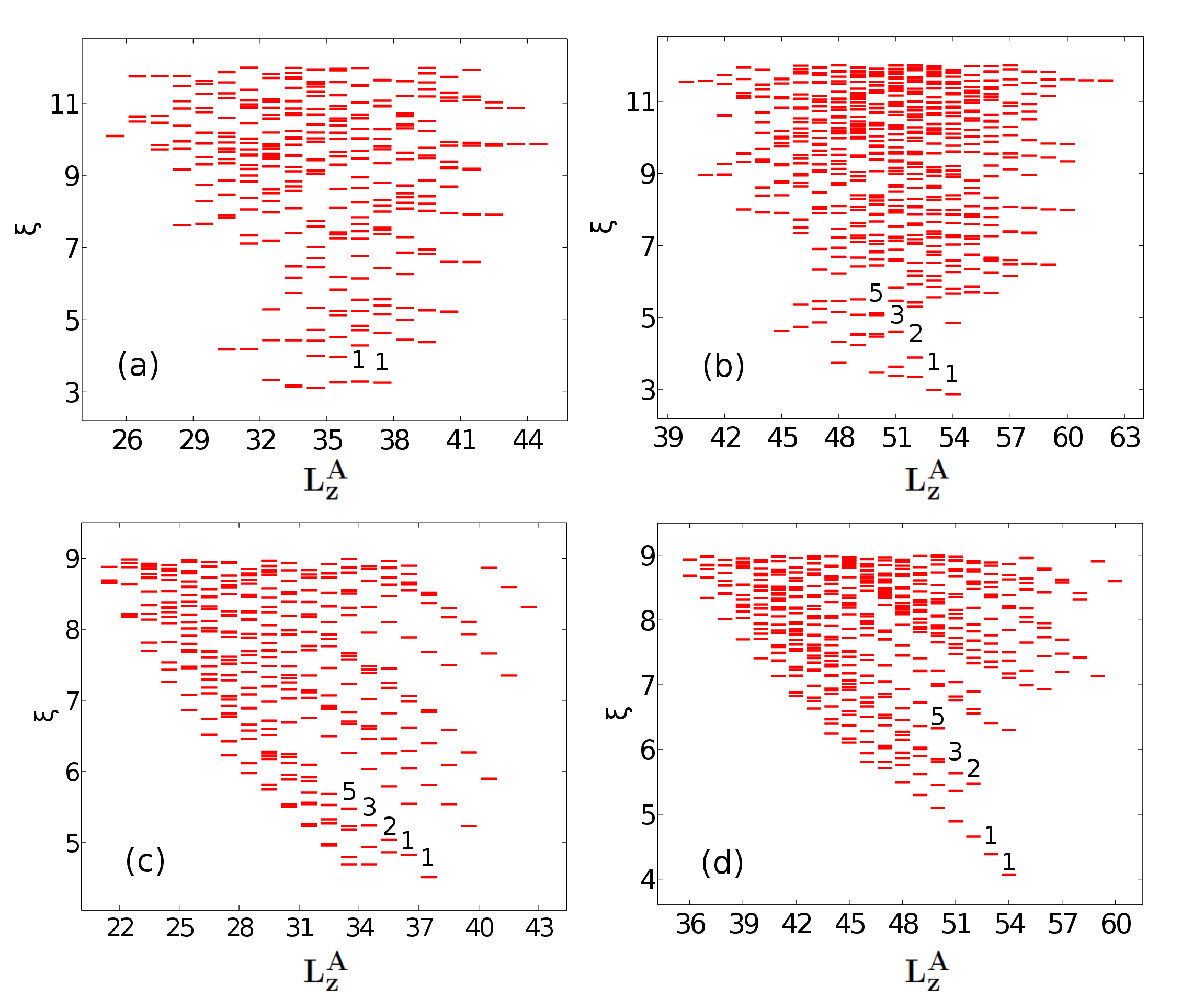}
\caption{(color online) Entanglement spectra of the exact Coulomb 7/3 state for 10 and 12 electrons. Orbital entanglement spectrum for (a)  $N=10$, $N_A=5$ and $l_A=14$ and (b) $N=12$, $N_A=6$ and $l_A=17$. Real space entanglement spectra for (c) $N=10$ and $N_A=5$ and (d) $N=12$ and $N_A=6$. The counting of relevant states is indicated.}
\label{ES}
\end{figure}

{\em Entanglement spectrum (ES)}: We next study edge excitations by calculating the 
ES,\cite{Li08,Sterdyniak11,Dubail12,Sterdyniak12,Rodriguez12} which has been shown to capture 
the structure of the FQHE edge. The orbital ES \cite{Li08} (OES) and real space ES \cite{Dubail12,Sterdyniak12,Rodriguez12} (RSES) of Laughlin state show the counting of $1, 1, 2, 3, 5, \ldots$ as we move away from the maximum angular momentum, consistent with an edge described in terms of a single boson.\cite{Wen} 
Not surprisingly, the Coulomb $1/3$ state also shows the same counting. The situation is more subtle for 7/3, however. 
To calculate the ES, we cut the Hilbert space into two parts $A$ and $B$ and then calculate the reduced density matrix of $A$ by a partial trace of the density matrix over the degrees of freedom in $B$, {\em i.e.} $\widehat{\rho}_A={\rm Tr}_B |\Psi\rangle\langle\Psi|$, where $\Psi$ is the ground state. The eigenvalues of $\widehat{\rho}_A$ have the form $e^{-\xi}$ and a plot of $\xi$ produces the ES. In the OES, 
the region $A$ contains the $l_A$ orbitals localized in the northern hemisphere. In the RSES, the real space is partitioned and we choose the $A$ region to be the northern hemisphere. The OES is closely related to the RSES \cite{Dubail12,Sterdyniak12} since the single-particle orbitals are localized in real space. In both cases, $\widehat{\rho}_A$ commutes with $\widehat{N}_A$ and $\widehat{L}^A_z$, the particle number operator and the $z$ component of the orbital angular momentum operator of region $A$, so $\widehat{\rho}_A$ can be reduced to blocks labeled by $N_A$ and $L^A_z$. As shown in panels (a) and (b) of Fig.~\ref{ES}, the OES for the $N=10$ case does not have the single boson edge structure, but we can identify a counting of $1, 1, 2, 3, 5$ in the $N=12$ spectrum,
which suggests that the single boson edge spectrum is emerging with increasing $N$. For the RSES 
the single boson edge spectrum counting can be observed for both $N=10$ and $N=12$ as shown in panels (c) and (d) of Fig.~\ref{ES},  demonstrating that the RSES better captures the thermodynamic behavior than the OES.

\begin{figure}
\includegraphics[width=0.45\textwidth, height=0.28\textwidth]{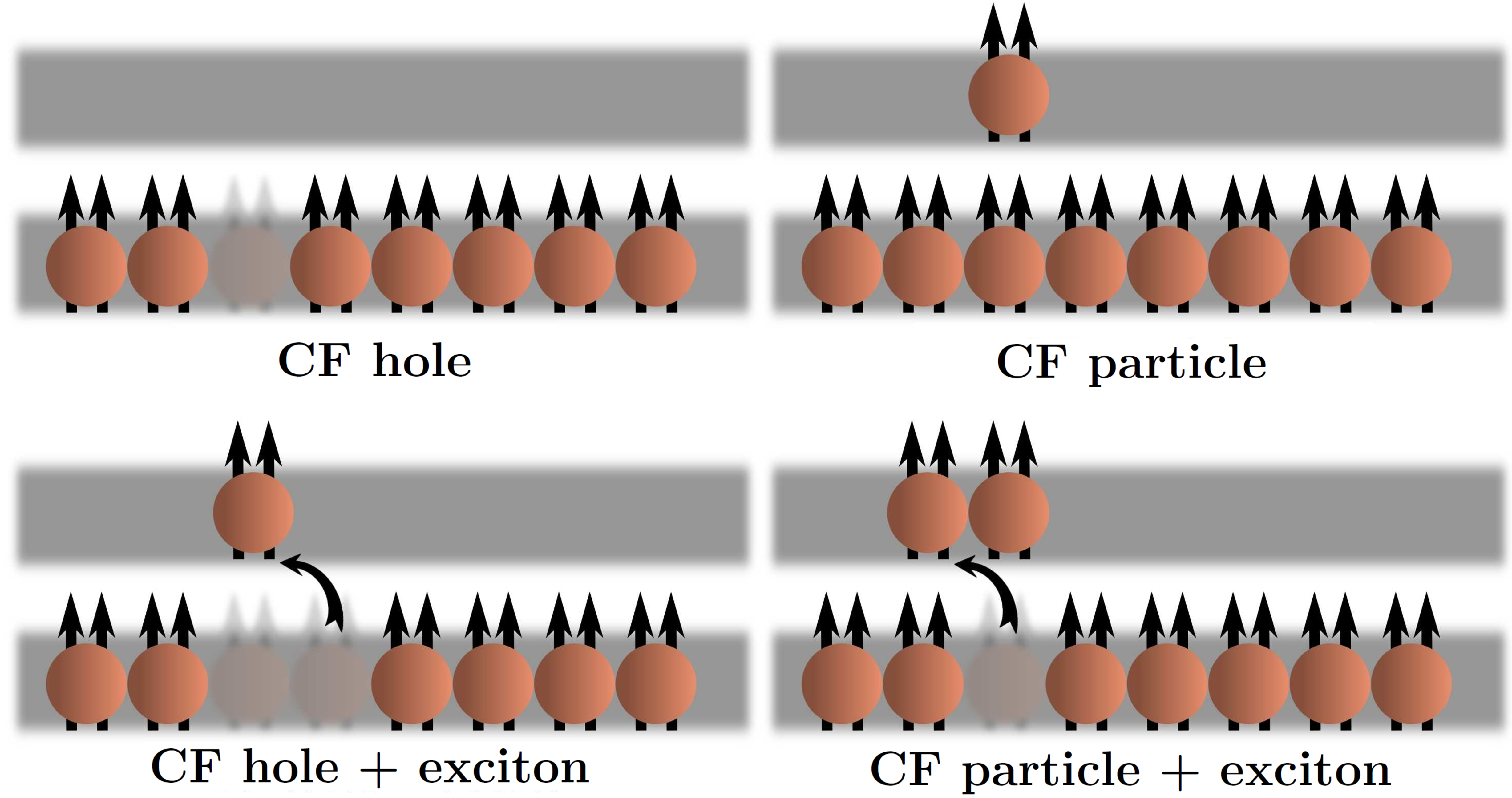}
\caption{(color online) 
The ``bare" CF hole at 1/3 is a missing composite fermion in the lowest $\Lambda$ level ($\Lambda$L) (upper left), while the ``bare" CF particle is an additional composite fermion in the second $\Lambda$L (upper right). The lower panels show schematically a ``dressed" CF hole and a ``dressed" CF particle screened by a single exciton, where the exciton is a neutral CF particle-hole pair. The horizontal lines are $\Lambda$Ls, and the spheres with two arrows represent composite fermions carrying two vortices.}
\label{fig:ext}
\end{figure}

{\em CF-exciton screening}:  The fact that in both the lowest and the second LLs, Coulomb interaction produces incompressible states (i.e. uniform $L=0$ ground states with a gap) at $2Q=3N-3$ with similar entanglement spectra suggests similarity between the FQHEs at 7/3 and 1/3. Yet, the exact Coulomb QP and QH excitations are strikingly different for these two states. How can we reconcile these apparently conflicting messages? In what follows, we show that a possible resolution to this puzzle lies in a strong screening by CF excitons\cite{Girvin,Scarola} at 7/3. One may expect such screening to be significant at 7/3 because here, compared to 1/3, the exciton gap is much smaller and the residual interaction between composite fermions are much stronger.

\begin{figure}
\includegraphics[width=0.45\textwidth]{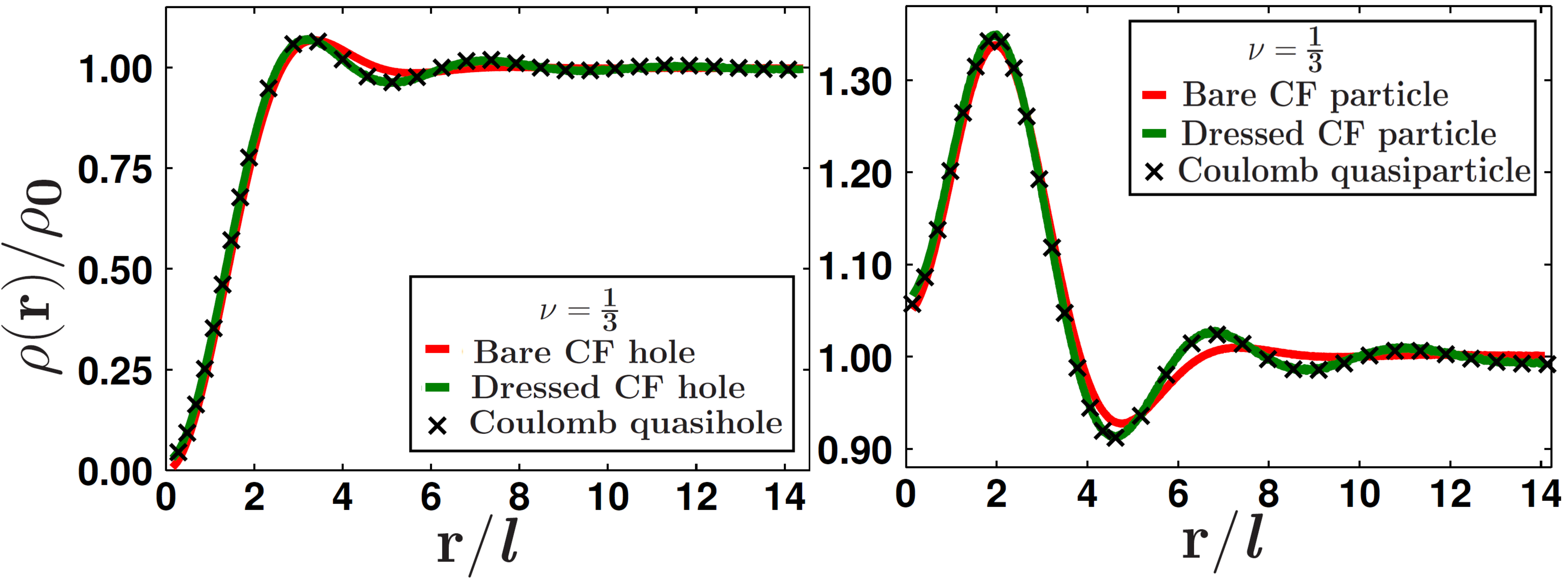}
\caption{(color online) The density profiles $\rho(r)/\rho_0$ at $\nu=1/3$ for bare and dressed CF hole and CF particle, along with those for the exact Coulomb quasiparticle and quasihole, located at the origin (the North Pole). Here $\rho_0$ is the density of the background FQHE state. The results are for a system of $N=15$ particles.}
\label{fig13}
\end{figure}

\begin{figure}
\includegraphics[width=0.45\textwidth]{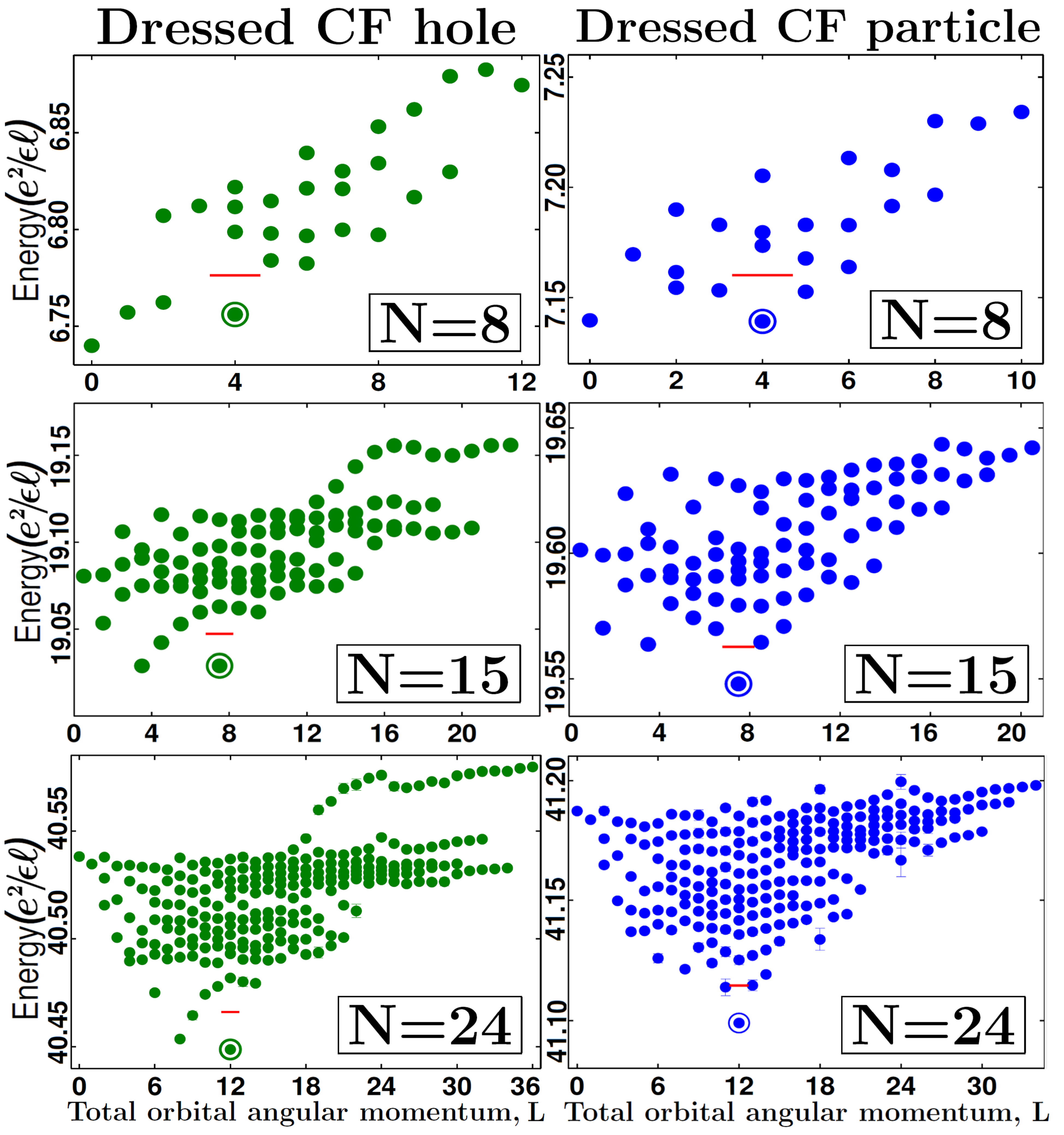}
\caption{(color online) Spectra for CF hole (left) and CF particle (right) at $\nu=7/3$ obtained from CF diagonalization.  The dressed CF hole and the dressed CF particle are encircled.  The energy of the bare CFH / CFP is shown by a red dash.}
\label{fig4}
\end{figure}

We recall the basic facts of the CF theory.\cite{Jain89} Composite fermions are bound states of electrons and quantized vortices. They experience an effective magnetic field, and form LL-like levels called $\Lambda$ levels ($\Lambda$Ls). Their filling $\nu^*$ is related to the electron filling factor $\nu$ by the relation $\nu=\nu^*/(2p\nu^*\pm 1)$. In particular, the filling $\nu=1/3$ maps into $\nu^*=1$ of composite fermions. The ground state at 1/3 corresponds to Laughlin's wave function. The CF hole (CFH) and CF particle (CFP) are shown schematically in the upper panels of Fig.~\ref{fig:ext}. The ``dressed" CFP  and CFH are shown schematically in the lower panels of Fig.~\ref{fig:ext}. To estimate the quantitative effect of screening, we perform CF diagonalization\cite{Mandal02} (CFD) in the extended subspace containing the states shown in both upper and lower panels of Fig.~\ref{fig:ext}.

For completeness, we give an outline of the CFD procedure.\cite{Mandal02} We first construct a basis \{$\Phi_{\alpha}^{\rm (1)}$\} of all Slater determinant states at $\nu^*=1$ of the type shown in both panels of Fig.~\ref{fig:ext}. These include a single CFH / CFP {\em and} a CFH / CFP $+$ a CF particle hole pair. We diagonalize $L^2$ to obtain the basis in each angular momentum sector, denoted \{$\Phi_{L,\alpha}^{\rm (1)}$\}. We then composite-fermionize \cite{Jain89} this basis according to \{$\Psi_{L,\alpha}^{\rm (1)}=\mathcal{P}_{\text{LLL}}\prod_{j<k}(u_jv_k-u_kv_j)^{2}\Phi_{L,\alpha}^{\rm (1)}$\},
where $\mathcal{P}_{\text{LLL}}$ denotes projection into the LLL, $\alpha$ labels different states at a given $L$, and $u=\cos(\theta/2)e^{i\frac{\phi}{2}}$ and $v=\sin(\theta/2)e^{-i\frac{\phi}{2}}$ are spinor coordinates of a particle on a sphere. (The $L$ quantum number is conserved under this mapping.) The LLL projection is carried out by the method in Ref.~\onlinecite{JK}. We finally obtain the spectrum by diagonalizing within the basis \{$\Psi_{L,\alpha}^{\rm (1)}$\} the Coulomb Hamiltonian. The orthogonalization of the basis functions and the determination of the interaction matrix elements require evaluation of multidimensional integrals, which is performed using the Monte Carlo method described previously.\cite{Mandal02}

\begin{figure}
\includegraphics[width=0.45\textwidth]{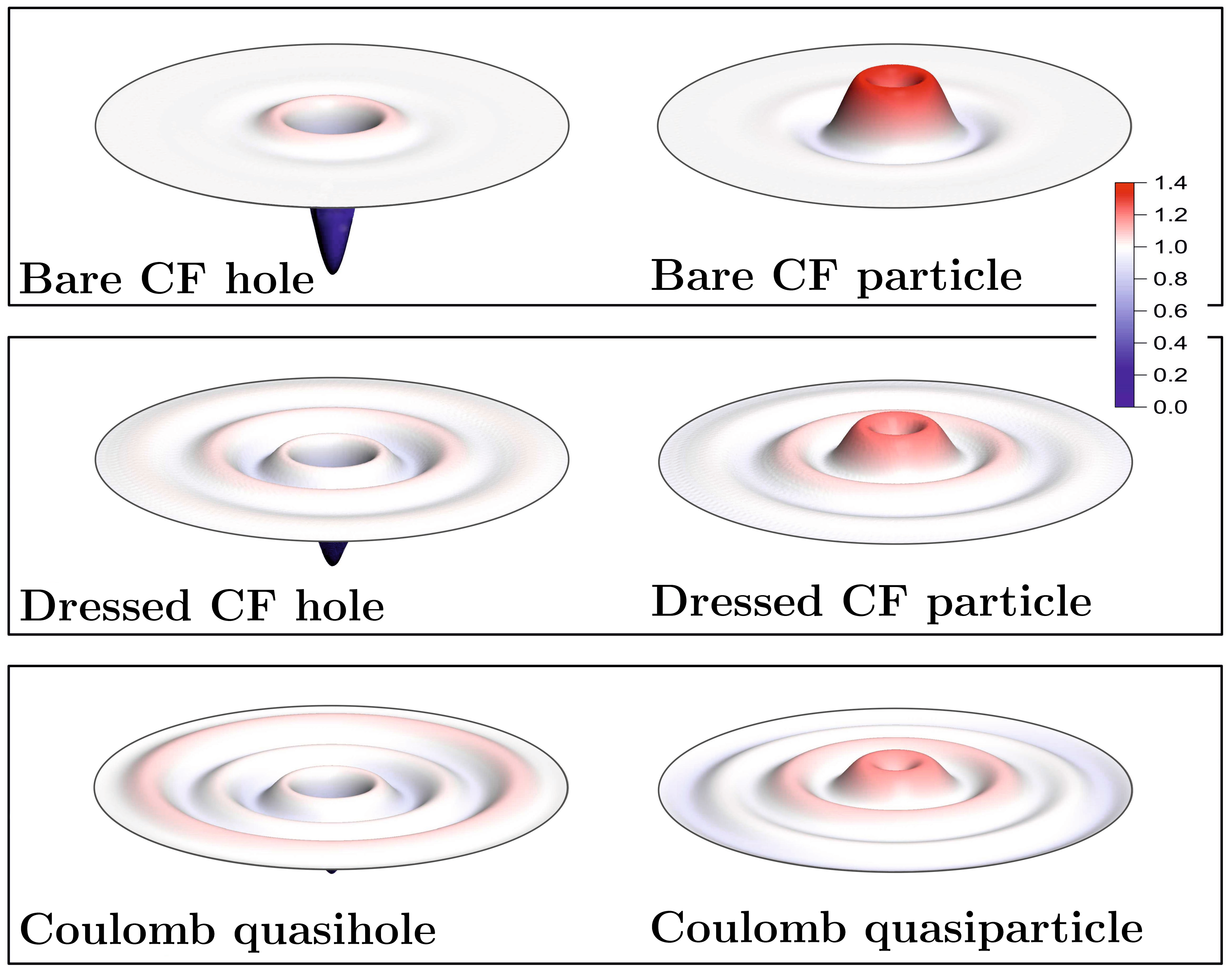}
\caption{(color online) The density profiles $\rho(r)/\rho_0$ for bare and dressed CF hole and CF particle, along with those for the exact Coulomb quasihole and quasiparticle, located at the origin (the North Pole). Here $\rho_0$ is the density of the background FQHE state, and LLL orbitals are used to obtain all densities. The results are for a system of $N=15$ particles at $\nu=7/3$, and the diameter of the disk shown is $\sim$28 $\ell$. The bare CFP and CFH are very close to the Coulomb ones at $\nu=1/3$.}
\label{fig:density_QPs}
\end{figure}

We begin by studying the effect of screening by CF excitons at 1/3. Here, the bare CFH and CFP are already very accurate, as shown in Fig.~\ref{fig13}. Screening by excitons makes only a small difference, but brings the dressed CFP and CFH into an almost exact agreement with the actual Coulomb QP and QH. 
The situation at 7/3 is very different. The 7/3 state is simulated in the LLL by working with the interaction 
\begin{equation}
V^{\text{eff}}(r)=\frac{1}{r}+\frac{B_3}{\sqrt{r^6+1}}+\frac{B_5}{\sqrt{r^{10}+10}}+ (C_0+C_1r^2+C_2r^4)e^{-r^2} \nonumber
\end{equation}
where $r$ is in units of magnetic length $\ell$. With $B_3=1$, $B_5=2.25$, $C_0=-20.94019$, $C_1=12.98214$ and $C_2=-1.53215$, this interaction produces almost exactly the same matrix elements in the LLL as the Coulomb interaction in the second LL.\cite{Chuntai,Toke_Jain_06} 
The CFD spectra are shown in Fig.~\ref{fig4}; red dashes mark the energy of the {\em bare} CFP / CFH. We will tentatively identify the lowest $L=N/2$ state as the dressed QP / QH. A substantial dressing of the CFP / CFH by excitons is indicated by a lowering of its energy by an amount comparable to the 7/3 gap,\cite{Wojs09} and also by a significant change in the density profiles (Fig.~\ref{fig:density_QPs}), which shows that the the dressed CFP and CFH have a much larger size than the bare ones. This figure also shows the density profiles of the Coulomb QP and QH, which have a diameter of $> 30$ $\ell$, much larger than $\sim$ 12 $\ell$ for the bare particles. While the density of the dressed CFP and CFH at 7/3 agrees well with the density of the Coulomb QP and QH at small $r$, deviations appear at larger $r$, implying that dressing by a {\em single} exciton is inadequate. Inclusion of two or more excitons will bring the dressed particles closer to the real ones, but we have not pursued that.

The large QP / QH size at 7/3 provides a possible rationalization for the puzzling behavior seen in exact diagonalization studies (e.g. Fig.~1): these are unable to capture the true character of the QP or QH because the system sizes are small compared to the size of the QP or QH.  This view is supported by the $N$ dependence of CFD spectra. The state at $L= N/2$ (marked by a circle in Fig. \ref{fig4}) does not have the lowest energy in small systems, but with increasing $N$, it  moves down, becoming the lowest energy state for sufficiently large $N$. The value of $N$ where the crossover takes place for the QP does not match precisely in exact and CF diagonalizations,\cite{cross} which is not surprising, because (i) our interaction accurately reproduces the second LL pseudopotentials in the planar geometry but will have finite size corrections on the sphere, and (ii) screening by more than one exciton has been neglected. Nonetheless, we believe that our CFD results combined with the ground state studies and entanglement spectra make it plausible that for sufficiently large $N$ the lowest energy Coulomb states may also occur at $L=N/2$. That makes it likely that 7/3 and 1/3 are governed by similar physics. In other words, while the exciton screening has a large quantitative effect at 7/3, it probably does not cause a phase transition.

The above physics is somewhat reminiscent of skyrmions, wherein the spin texture of a spin reversed electron at $\nu=1$ is enlarged due to dressing by low energy spin waves.\cite{Sondhi} The study of that complex excitation has given rise to much elegant physics. We do not consider the spin degree of freedom above.

{\em Implications for experiments}: 
The renormalization of the 7/3 CFP and CFH implies a renormalization of composite fermions in the 7/3 ground state as well. In other words, the 7/3 state is to be thought of as one filled $\Lambda$L of {\em dressed} composite fermions. That explains why it is poorly described in terms of free composite fermions, and why it is much less robust than the 1/3 state. One may ask if other fractions of the form $2+n/(2n\pm 1)$ can be understood as filled $\Lambda$Ls of dressed composite fermions; numerical studies\cite{Sreejith} do not support a two filled $\Lambda$L description of $2+2/5$ for systems with zero width. We note that screening by excitons is, of course, always present -- it is fortunate that such screening has a negligible effect in the LLL, allowing an accurate quantitative description in terms of bare composite fermions.

The strikingly large size of the QP and QH ($\sim$ 500 nm for typical experimental conditions) has relevance to several experiments. A substantial enhancement in tunneling across a narrow constriction at 7/3, as compared to 1/3 (or 5/3), may be observable. The large size of the 7/3 QPs / QHs should make an interference measurement of their braid statistics\cite{Halperin} more challenging, which requires that the braiding particles do not overlap.\cite{Jeon} It is also relevant to the issue of localization of composite fermions by disorder, because the localization length is sensitively dependent on the size of the localized particle. These considerations are pertinent to the local electrometry measurements,\cite{Venkatachalam} Aharanov Bohm oscillations,\cite{Willett} and interference spectroscopy \cite{Kang} where phase slips have been observed and interpreted in terms of fractional braid statistics. The large size of dressed CFP / CFH implies that their crystal in the vicinity of 7/3 is unlikely; evidence for such a crystal in the vicinity of 1/3 is seen in microwave experiments,\cite{Zhu1} which show conductivity resonances interpreted as pinning modes of the CFP and CFH crystals.\cite{Zhu1,CFWC} Finally, the complex structure of the 7/3 QP and QH should make them more susceptible to external perturbations, such as a mass anisotropy.\cite{Eisenstein,Shayegan}

We acknowledge financial support from the NSF grant no. DMR-1005536 (ACB), the DOE Grant No. DE-SC0005042 (YHW and GJS), and the Polish NCN grant 2011/01/B/ST3/04504 and EU Marie Curie Grant PCIG09-GA-2011-294186 (AW). We thank Research Computing and Cyberinfrastructure, a unit of Information Technology Services at Pennsylvania State University, as well as Wroclaw Centre for Networking and Supercomputing and Academic Computer Centre CYFRONET, both parts of PL-Grid Infrastructure.


\begin{thebibliography}{99}
\bibitem{comm} The terms QP and QH will refer exclusively to the exact Coulomb quasiparticle and quasihole.
\bibitem{cross} For QP, the crossover does not occur until $N=15$ in exact diagonalization, but occurs already at $N=10$ in CFD.
\end{thebibliography}
\end{document}